\documentclass[12pt]{article}

\usepackage{scicite}

\usepackage{times}

\newenvironment{sciabstract}{%
\begin{quote} \bf}
{\end{quote}}

\usepackage{graphicx} % For initial submission

\usepackage[version=3]{mhchem} % Formula subscripts using \ce{}

\usepackage{tipa} % For latin gamma symbol
\newcommand{\gammalatin}{\text{\textipa{G}}} 

\usepackage{amssymb} % For  \gtrsim

\renewcommand\vec{\mathbf} % Bold font for vectors

\usepackage{braket} % For bra-ket notation

\title{Twist Angle Tuning of Moir\'{e} Exciton Polaritons in van der Waals Heterostructures}

\author
{Jamie M. Fitzgerald,$^{1\ast}$ Joshua J. P. Thompson,$^{2}$ Ermin Malic$^{1,2}$\\
\\
\normalsize{$^{1}$Department of Physics, Chalmers University of Technology, SE-412 96 Gothenburg, Sweden}\\
\normalsize{$^{2}$Fachbereich Physik, Philipps-Universität, Marburg, 35032, Germany}\\
\normalsize{$^\ast$To whom correspondence should be addressed; E-mail:  jamief@chalmers.se}
}

\date{}

\begin{document} 


\baselineskip24pt


\maketitle


\begin{sciabstract}
Twisted atomically thin semiconductors are characterized by moir\'{e} excitons. Their optical signatures and selection rules are well understood. However, their hybridization with photons in the strong coupling regime for heterostructures integrated in an optical cavity has not been in the focus of research yet. Here, we combine an excitonic density matrix formalism with a Hopfield approach to provide microscopic insights into moir\'{e} exciton polaritons. In particular, we show that exciton-light coupling, polariton energy, and even the number of polariton branches can be controlled via the twist angle. We find that these new hybrid light-exciton states become delocalized relative to the constituent excitons due to the mixing with light and higher-energy excitons. The system can be interpreted as a natural quantum metamaterial with a periodicity that can be engineered via the twist angle. Our study presents a significant advance in microscopic understanding and control of moir\'{e} exciton polaritons in twisted atomically thin semiconductors. 
\end{sciabstract}

% In setting up this template for *Science* papers, we've used both
% the \section* command and the \paragraph* command for topical
% divisions.  Which you use will of course depend on the type of paper
% you're writing.  Review Articles tend to have displayed headings, for
% which \section* is more appropriate; Research Articles, when they have
% formal topical divisions at all, tend to signal them with bold text
% that runs into the paragraph, for which \paragraph* is the right
% choice.  Either way, use the asterisk (*) modifier, as shown, to

\section*{Introduction}

Two monolayers of transition metal dichalcogenides (TMDs) can be vertically stacked to form a type-II heterostructure \cite{rivera2015observation, kunstmann2018momentum, liu2019recent, merkl2020twist}. \ce{MoSe2}/\ce{WSe2} is a typical example, which exhibits a large band offset such that electronic hybridisation at the K point is negligibly small \cite{gillen2018interlayer}, leading to well-defined intralayer and interlayer exciton states \cite{ovesen2019interlayer}. The former have a large oscillator strength and manifest as a visible signal in absorption spectra. The latter are spatially indirect, exhibit ultra-long lifetimes, and were shown to dominate low-temperature photoluminescence spectra despite their small oscillator strength  \cite{nagler2017interlayer,ovesen2019interlayer}. It has also been demonstrated that artificial moir\'{e} superlattices can be formed by engineering a finite twist angle between TMD layers \cite{yu2017moire, jin2019observation, tran2019evidence}. One consequence is the emergence of multiple flat exciton minibands significantly modifying optical emission \cite{seyler2019signatures, alexeev2019resonantly} and absorption spectra \cite{zhang2018moire,jin2019observation, brem2020tunable,forg2021moire} of TMD heterobilayers. In a twisted geometry, the electronic band energies at the K point vary periodically in space reflecting the local atomic registry of neighbouring layers. If the resulting moir\'{e} potential is deep enough, low-energy excitons can be trapped and hopping between adjacent supercells strongly suppressed \cite{wu2017topological,wu2018theory,brem2020tunable, baek2020highly, choi2020moire}.

Excitons in monolayer TMDs have a large oscillator strength and binding energy, making them suitable for integration into optical microcavities and for exploration of the strong-coupling regime \cite{schneider2018two}. Here, the light-matter coupling strength exceeds dissipation in the material and radiative decay from the cavity \cite{kavokin2003thin}. In this regime, the formation of exciton polaritons, which are hybrid light-exciton quasi-particles, has been explored theoretically \cite{vasilevskiy2015exciton,gutierrez2018polariton,latini2019cavity} and observed experimentally for TMDs placed in conventional dielectric Fabry-Perot cavities \cite{liu2015strong, dufferwiel2015exciton}, as well as using Tamm-plasmon photonic microstructures \cite{lundt2016room}, sub-wavelength-thick photonic crystals \cite{zhang2018photonic}, plasmonic lattices \cite{liu2016strong}, and nanodisks \cite{verre2019transition}. Unique properties of TMDs have been exploited to demonstrate valley polarized polaritons \cite{sun2017optical,dufferwiel2018valley} and trion polaritons \cite{emmanuele2020highly}. The study of exciton polaritons has led to a wealth of interesting phenomena in fundamental physics, such as Bose-Einstein condensation \cite{kasprzak2006bose, anton2021bosonic} and superfluidity \cite{amo2009superfluidity}, as well as practical applications such as polaritonic lasing \cite{kang2019room}.

Little is known about the strong-coupling physics of twisted heterostructures and the impact of moir\'{e} superlattices. In particular, this is relevant for exploring the collective light-matter coupling in arrays of quantum particles, representing a solid state analogue to atomic optical lattices \cite{ritsch2013cold}. Recently, there was a first experimental demonstration of polaritons in a twisted \ce{WS2}/\ce{MoSe2} heterobilayer \cite{zhang2021van}, where the density dependence of the localized moir\'{e} polaritons revealed a strong non-linearity due to the exciton blockade. For \ce{MoSe2}/\ce{WSe2} heterobilayers specifically, only the weak-coupling regime has been explored so far, demonstrating the Purcell enhancement of the interlayer exciton light-matter interaction \cite{forg2019cavity}, and interlayer exciton lasing \cite{paik2019interlayer}. 

In this work, we develop a microscopic model of moir\'{e} exciton polaritons. In particular, we focus on the twist angle as a new knob to control their optical response. We study the strong coupling between \emph{intralayer} moir\'{e} excitons and cavity photons in a twisted AA-stacked \ce{MoSe2}/\ce{WSe2} heterobilayer placed in the centre of a Fabry-Perot cavity, cf. Fig.~\ref{fig:fig1}(a). This exemplary heterostructure has been the focus of recent studies \cite{seyler2019signatures,brem2020tunable,yu2020electrically}, and the obtained qualitative insights are applicable also to other TMD-based heterobilayers. We use an excitonic density matrix formalism with a small-angle effective continuum model to describe moir\'{e} excitons. The interaction with cavity modes is included by performing a Hopfield diagonalization \cite{hopfield1958theory}, revealing multiple distinct branches of hybrid exciton polaritons. Specifically, we find that the mixing with light and higher-energy delocalized excitons inevitably leads to a cavity-controllable delocalization of moir\'{e} exciton polaritons. The polariton energy strongly depends on the twist angle due to: (i) an inherited dependence from the constituent excitons, and (ii) a twist-dependent Rabi splitting from the distribution of oscillator strength amongst the excitons. We also find that controlling the exciton-light interaction via the cavity length can drastically change the nature of the twist-angle dependence of different polariton branches, highlighting the dual light-matter character and control of moir\'{e} exciton polaritons.

%%%%%%%%%%%%%
%% Theory %%
%%%%%%%%%%%%%
\section*{Theoretical approach}
The moir\'{e} exciton Hamiltonian is modelled within the tight-binding approximation by assuming that the moir\'{e} potential stems from the interaction of the neighbouring layer's d-orbitals, which are the major orbital contribution at the K points \cite{brem2020tunable,wang2017interlayer}. The model is valid in the continuum limit of small angles \cite{dos2012continuum, wu2017topological,wu2018theory}, i.e. large moir\'{e} supercell periods, which is invaluable as this is an extremely challenging regime for first-principle calculations. To model the excitonic behaviour, we first solve the bilayer Wannier equation for a given layer configuration of the constituent electron and hole, taking into account the screening effect of the immediate dielectric environment \cite{ovesen2019interlayer}. This allows for a transformation into an effective single-particle Hamiltonian using an exciton basis, $\hat{H}=\sum_{\vec{Q}}E^{(X)}_{\vec{Q}}\hat{X}_{\vec{Q}}^\dagger\hat{X}_{\vec{Q}}+\sum_{\vec{Q},\vec{q}}\mathcal{M}_{\vec{q}}\hat{X}_{\vec{Q}+\vec{q}}^\dagger\hat{X}_{\vec{Q}}$, where $\hat{X}_{\vec{Q}}$ is the exciton field operator at the center-of-mass momentum $\vec{Q}$, and $\mathcal{M}_{\vec{q}}$ the matrix dictating the mixing of excitons by the moir\'{e} potential. Next, a zone-folding approach is applied, where $\vec{Q}$ is restricted to the mini Brillouin zone (mBZ). A projection to a moir\'{e} exciton basis gives \cite{brem2020tunable}, $\hat{H} = \sum_{\nu,\vec{Q}} E_{\nu}^{(Y)}(\vec{Q}) \hat{Y}_{\nu,\vec{Q}}^\dagger\hat{Y}_{\nu,\vec{Q}}
$, with $\vec{Q} \in \text{mBZ}$ and $\hat{Y}_{\nu,\vec{Q}} = \sum_{i,j} \mathcal{C}_{\nu,i,j}(\vec{Q}) \hat{X}_{\vec{Q}+i\vec{g}_0 + j \vec{g}_1}$, where the vectors $\vec{g}_0$ and $\vec{g}_1$ span the mBZ. The moir\'{e} exciton wavefunction and energy are given by $\mathcal{C}_{\nu,i,j}(\vec{Q})$ and $E_{\nu}^{(Y)}(\vec{Q})$, respectively. 

The bandstructure for the \ce{MoSe2}-based intralayer exciton at a twist angle of $1^\circ$ is shown in Fig.~\ref{fig:fig1}(b). Below the unperturbed 1s intralayer exciton energy (i.e. neglecting moir\'{e} effects and denoted by the black-dashed line), three flat bands are found, indicating that these excitons are localised within the moir\'{e} potential. Excitons at the $\gamma$ point (centre of the mBZ) are located within the light cone and can interact with light. However, due to symmetry reasons some of these branches interact very weakly with light, behaving as dark excitons. This is dictated by the value of the wavefunction at the $\gamma$ point, $\mathcal{C}_{\nu,i=j=0}(0)$. The radiative coupling of the $\nu$th exciton is given by $\hbar \gammalatin_\nu^{(Y)} = \hbar \gammalatin^{(X)} |\mathcal{C}_{\nu,i=j=0}(0)|^2$, where $\hbar \gammalatin^{(X)}$ is the coupling in the absence of moir\'{e} effects (see Methods and supplementary text). This is analogous to the regular expression for the radiative coupling of a monolayer, which depends on the value of the exciton wavefunction at zero relative momentum \cite{kira2006many}. The radiative coupling is shared between moir\'{e} excitons at the $\gamma$ point of the mBZ, with the total oscillator strength conserved. It shows a twist-angle dependence via the moir\'{e} wavefunctions. The different exciton sub-bands are color coded by the magnitude of the oscillator strength at the mBZ centre, cf. Fig. \ref{fig:fig1}(b). The bands labelled $\mathcal{Y}_1$ and $\mathcal{Y}_2$ correspond to the two energetically lowest bright excitons. As we are interested in absorption spectra, dark excitons are not further considered.

Placing the heterobilayer in a cavity leads to an enhanced resonant coupling between moir\'{e} excitons and cavity photons. The corresponding Hamiltonian in a rotating-wave approximation reads
\begin{equation}
    \hat{H} = \sum_{\nu=1}^N E_{\nu}^{(Y)}\hat{Y}_{\nu}^\dagger\hat{Y}_{\nu}
+ \hbar \omega \hat{b}^\dagger \hat{b} 
+ \sum_{\nu=1}^N g_\nu \left( \hat{Y}_{\nu} \hat{b}^\dagger + \hat{Y}^\dagger_{\nu} \hat{b} \right),\label{eq:LM_Hamiltonian}
\end{equation}
and can be converted to the exciton-polariton basis via a Hopfield transformation \cite{hopfield1958theory}. The $n$th-band polariton field operator is expressed as
$\hat{\mathcal{P}}_n = U_{0n} \hat{b}  + \sum_{m=1}^N U_{mn} \hat{Y}_m,$
where $\hat{b}$ is the cavity-photon field operator. The expansion coefficients $U_{mn}$ are the eigenvectors obtained from diagonalising Eq.~(\ref{eq:LM_Hamiltonian}) and are known as Hopfield coefficients. They measure the contributions of the $m$th constituent photon/exciton to the nth polariton state. In particular, $U_{0n}$ quantifies the photonic nature of the nth polariton. An advantage of this method is that we can derive intuitive expressions for both polariton absorption and effective mass in terms of Hopfield coefficients, in contrast to the alternative classical T-matrix method \cite{kavokin2003thin}. For smaller cavities, we can consider only a single cavity mode due to the large free-spectral range relative to the exciton energy spacing. This means that for $N$ excitons, there are $N+1$ polaritons. The coupling can be determined analytically for the specific case of a thin excitonic material placed in the centre of a high-quality symmetric Fabry-Perot cavity at frequencies close to the cavity resonance \cite{kavokin2003thin}
\begin{equation}
g_{\nu} = \hbar \sqrt{\frac{1+|r_m|}{|r_m|}\frac{ \gammalatin_\nu^{(Y)}}{\tau}} \approx  \sqrt{ \frac{2}{\pi} \hbar \gammalatin_\nu^{(Y)} E_\nu^{(Y)}}, \label{eq:coupling}
\end{equation}
where $r_m$ is the end-mirror reflectivity, $\tau=L /c$ the one-way photon travel time in the cavity, and $L$ the cavity length. Note that the coupling will obtain a twist-angle dependence via the radiative recombination. The limit holds for a perfect ($|r_m|\approx 1$) cavity tuned near the $\nu$th exciton energy. For typical \ce{MoSe2} parameters \cite{gillen2021interlayer}, the coupling magnitude will be about $40$ meV for the intralayer exciton in the absence of moir\'{e} effects.

\section*{Results}

\subsection*{Cavity Control of Moir\'{e} Exciton Polaritons}

\begin{figure}[t!]
\includegraphics[width=\columnwidth*3/4]{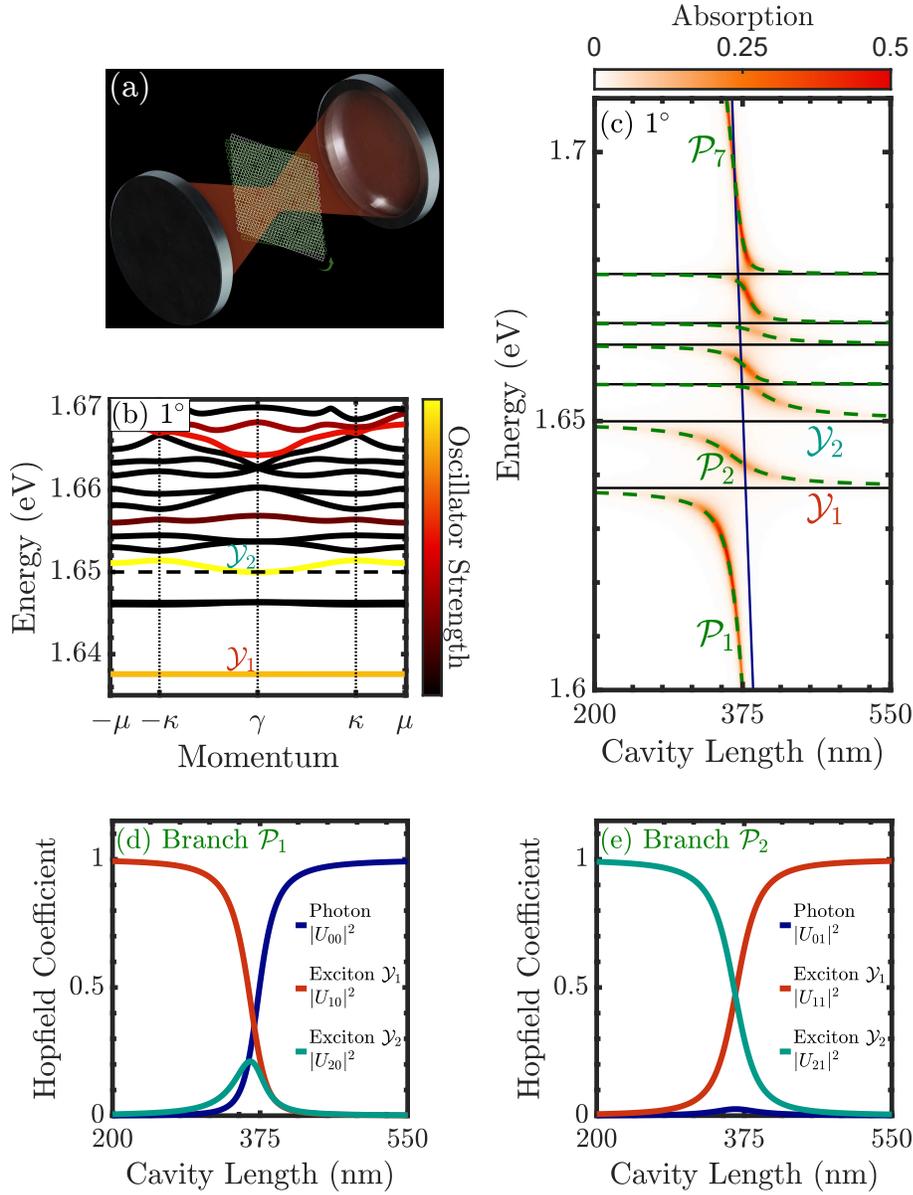}
\caption{\label{fig:fig1} (a)~Schematic illustration of a \ce{MoSe2}/\ce{WSe2} heterobilayer placed in the centre of an Fabry-Perot cavity. (b)~Calculated moir\'{e} exciton minibands for the intralayer exciton at a $1^{\circ}$ twist angle. Colour coding is proportional to the oscillator strength. Black dashed line indicates the \ce{MoSe2}-based exciton energy in the absence of moir\'{e} effects. (c)~Cavity length dependence of moir\'{e} exciton polariton bands (dashed-green line) overlaid on the absorption for a $1^{\circ}$ twist angle. Bare exciton (cavity) energies are indicated by the black (blue) lines. Absolute square of the Hopfield coefficients, showing the photonic ($U_{0n}$) and excitonic contributions ($U_{1n}$ and $U_{2n}$) to the polariton branch (d)~$\mathcal{P}_1$ and (e)~$\mathcal{P}_2$. }
\end{figure}

One strategy to explore the strong-coupling regime is to detune photon and exciton energies by changing the cavity length. Here, we consider only cavity modes with a vanishing in-plane momentum. Figure \ref{fig:fig1}(c) shows the polariton energy as a function of cavity length for a fixed twist angle of $1^\circ$. Due to the large coupling of intralayer excitons to light, it is necessary to include both intralayer excitons located in the \ce{MoSe2} ($1.65$ eV in absence of moir\'{e} effects) and \ce{WSe2} layer ($1.75$ eV). There are also two interlayer configurations, in particular a low-energy interlayer exciton \cite{ovesen2019interlayer, brem2020tunable} at $1.35$ eV, but their small oscillator strength makes them challenging to utilize for strong-coupling physics. For simplicity we will first concentrate our analysis on \ce{MoSe2}-based intralayer excitons. The energy of the polariton branches is indicated by green-dashed lines in Fig.~\ref{fig:fig1}(c), and is overlaid on a colormap of the polariton absorption. A strong coupling between the cavity photon (blue line) and the six bright moir\'{e} excitons (flat black lines, corresponding to the exciton bands in Fig.~\ref{fig:fig1}(b) at the $\gamma$ point) is indicated by an avoided crossing of the resulting polariton dispersion near the points of intersection.  

The small energy separation between moir\'{e} exciton bands ($\sim 10$ meV at $1^\circ$) is on the same order of magnitude as the coupling strength between the individual moir\'{e} excitons and the cavity photon. This leads to a set of seven polariton branches including five middle branches of a mostly excitonic nature, and an upper and lower polariton branch with a strongly cavity length-dependent hybrid light-exciton nature. The lower polariton branch, labeled as $\mathcal{P}_1$, can be readily understood: for small cavity lengths ($L < \lambda/2$, where $\lambda$ is the cavity photon wavelength corresponding to the $\mathcal{Y}_1$ exciton energy), the cavity is off-resonant and the polariton branch follows the bare energy of the  $\mathcal{Y}_1$ exciton (black line). There is an avoided crossing near the intersection point around $378$ nm due to the hybridisation between the exciton and the photon. At longer cavity lengths, the polariton becomes increasingly light-like and follows the bare cavity mode (blue line). 

This inferred behaviour is confirmed by studying the Hopfield coefficients, cf. Fig.~\ref{fig:fig1}(d), where the contribution of the photon, and the energetically lowest $\mathcal{Y}_1$ and $\mathcal{Y}_2$ excitons is shown. They reveal that $\mathcal{P}_1$ also has a significant contribution from $\mathcal{Y}_2$ close to the avoided crossing. This photon-mediated exciton hybridisation occurs for polaritonic systems with multiple energetically closely-spaced excitons (relative to the coupling strength) \cite{lidzey2000photon}. Interestingly, the cavity can strongly mix excitons without the resulting polariton possessing a strong photonic component, as nicely illustrated by the $\mathcal{P}_2$ polariton in Fig.~\ref{fig:fig1}(e). Here, for smaller cavity lengths the polariton is predominately $\mathcal{Y}_2$ exciton-like, but as the cavity length is increased, the contribution of  the $\mathcal{Y}_1$ exciton grows. Around the avoided crossing, the polariton is an almost equal combination of the $\mathcal{Y}_1$ and the $\mathcal{Y}_2$ exciton, with only a small photonic contribution. This manifests as a flatter length dependence of the $\mathcal{P}_2$ polariton energy in this region (relative to $\mathcal{P}_1$), cf. Fig.~\ref{fig:fig1}(c). 

The polariton dispersion is not directly observable in experiments. It is therefore instructive to explore polariton absorption, which unambiguously demonstrates strong coupling via the Rabi splitting \cite{kavokin2017microcavities}. Absorption can be readily calculated using the T-matrix method \cite{kavokin2003thin}, but it is fruitful to combine the Hopfield approach with the input-output formalism \cite{collett1984squeezing}. In the limit of a small scattering loss, $\Gamma$, relative to energy spacing between excitons, we obtain an Elliot-like expression (see supplementary text)
\begin{equation}
A(\omega)= \sum_{n=0}^N \frac{4 \tilde{\gammalatin}_n  \tilde{\Gamma}_n}{( \omega- E_n^{(Y)}/\hbar)^2 + (2 \tilde{\gammalatin}_n + \tilde{\Gamma}_n)^2}
\label{eq:absorption} 
\end{equation}
with an effective radiative coupling $\tilde{\gammalatin}_n=c  T_m |U_{0n}|^2/4L$, and scattering loss $\tilde{\Gamma}_n=\Gamma\left(1-|U_{0n}|^2\right)$ for the $n$th polariton. The polaritonic Elliot formula is centred on the polariton energy $E_n^{(Y)}$ with a width determined by the sum of the effective non-radiative and radiative couplings. The latter plays the same role as the radiative coupling in the usual excitonic Elliot formula \cite{kira2006many}, and describes how polaritons couple to the external ports. Intuitively, it is equal to the bare-cavity decay rate, $\kappa=c T_m /4L$, scaled by the Hopfield coefficient describing the photonic contribution of the $n$th polariton, $|U_{0n}|^2$. The cavity properties enter directly through both the length, $L$, and the transmission of the end mirrors, $T_m$. The latter results from the fact that light may only couple to outside ports through the end mirrors. Similarly, the effective scattering loss is dependent on the total excitonic contribution to the polariton via the factor $1-|U_{0n}|^2=\sum_{\nu=1}^N |U_{\nu n}|^2$, where the latter relation is a consequence of Bose commutation relations for polaritons. This reflects that all non-radiative decay channels are included via excitons in our model. Equation (\ref{eq:absorption}) reveals that the absorption is dictated by a compromise between the excitonic and photonic contribution. In analogy to a bare excitonic system, the maximum possible absorption is $0.5$ due to the mirror symmetry of the system \cite{horng2020perfect} and occurs when total photonic decay (in this case through both ports) is equal to exciton scattering loss, i.e. $\tilde{\Gamma}_n=2\tilde{\gammalatin}_n$. The value of $|U_{0n}|^2$ at which this polaritonic critical-coupling condition is met is dictated by the balance between the exciton scattering rate and the cavity decay rate (see supplementary text). 

Taking the $\mathcal{P}_1$ polariton branch as an example, for small $L$ the cavity will not allow light in and no absorption can take place. This is captured by a vanishing $U_{0n}$, and hence $\tilde{\gammalatin}_n$, in Eq.~(\ref{eq:absorption}). For large $L$, $U_{0n}$ limits towards 1 and the contribution of the exciton scattering loss vanishes ($\tilde{\Gamma} \rightarrow 0$). Here, we recover the absorption of a bare cavity (which is zero in our model). For intermediate cavity lengths, close to the avoided crossings, we observe the largest absorption (cf. Fig.~\ref{fig:fig1}(c)). For the parameters used in this work ($\Gamma=1$ meV, $|r_m|=0.99$), peak absorption of $0.5$ is found at $|U_{0n}|^2=0.16$, which corresponds to a cavity length of $364$ nm (see Supplementary Fig.~(S1)). In contrast, the middle polariton branches, such as $\mathcal{P}_2$, are weakly photonic (small $|U_{0n}|^2$, see Fig.~\ref{fig:fig1}(e)) even near the avoided crossing region, and hence have a lower absorption compared to the  $\mathcal{P}_1$ branch. 

Finally, the analysis of the Hopfield coefficients for the higher-energy polariton branch $\mathcal{P}_7$ reveals that it predominately consists of the highest-energy moir\'{e} excitons from the \ce{MoSe2} layer and the lowest-energy excitons from the \ce{WSe2} layer, cf. Fig.~\ref{fig:fig1}(c). This represents an interesting example of photon-induced interlayer hybridisation and is similar to what has been previously reported for untwisted \ce{MoS2}/\ce{WS2} heterobilayers \cite{latini2019cavity}.

\subsection*{Localization of Moir\'{e} Exciton Polaritons}

A unique characteristic of low-energy moir\'{e} excitons is localization within the moir\'{e} potential for small twist angles \cite{wu2017topological,wu2018theory,brem2020tunable}. It is therefore interesting to study how this changes in the strong coupling regime. The cavity photon is completely delocalized in the transverse plane and the polariton is expected to at least partially inherit this property. One means of inferring the degree of localization is via the group velocity \cite{freixanet2000plane}, $v(\vec{k}_{\parallel})=\partial_{\vec{k}_{\parallel}} E(\vec{k}_{\parallel})/\hbar$, and the effective mass $m=\hbar^2/\partial_{\vec{k}_{\parallel}}^2E(\vec{k}_{\parallel})|_{\vec{k}_{\parallel}=0}$, where $\vec{k}_{\parallel}$ is the polariton in-plane momentum. For near-flat bands, i.e. the localized  $\mathcal{Y}_1$ exciton in Fig.~\ref{fig:fig1}(b), the group velocity vanishes over an extended region of the mBZ and the exciton branch is characterized by a large effective mass. As a result, hopping between moir\'{e} supercells is suppressed and the exciton acts as a large-mass effective free particle. 

\begin{figure}[t!]
\includegraphics[width=\columnwidth]{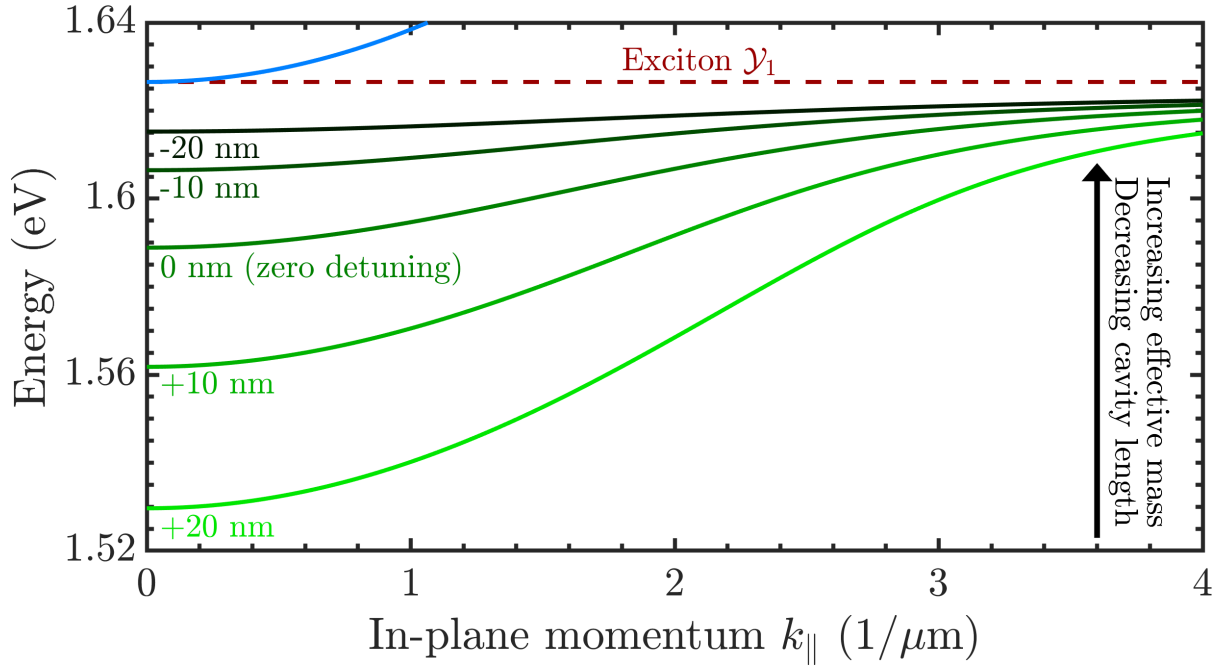}
\caption{\label{fig:fig2} Energy of the lowest polariton branch ($\mathcal{P}_1$) at a twist angle of $1^\circ$ as a function of in-plane momentum and for different cavity lengths (representing a detuning study). Exciton(cavity photon at zero detuning) dispersion is shown by the dashed-red(solid-blue) line for comparison. The curvature, and hence the group velocity and the effective mass of the polariton, are drastically altered by tuning the cavity resonance relative to the $\mathcal{Y}_1$ exciton energy.} 
\end{figure}

To this end, we perform a study for oblique cavity photons with a finite in-plane momentum, concentrating on the lowest-energy $\mathcal{Y}_1$ polariton branch for different detuning of the cavity photon and exciton energy at $\vec{k}_{\parallel}=0$, cf. Fig.~\ref{fig:fig2}. We focus on TM polarization, but very similar results are found for TE polarization around $\vec{k}_{\parallel}=0$. The large effective mass of the moir\'{e} exciton, $m^{(Y)}= 77 m_0$, relative to the effective photon mass, $m^{(c)}=E^{(c)}(\vec{k}_{\parallel}=0)/c^2 = 3\times 10^{-6}m_0$, means that the exciton branch is essentially flat (red dashed-line in Fig.~\ref{fig:fig2}) relative to the cavity dispersion $E^{(c)}=c\hbar\sqrt{ \pi^2/L^2 + |\vec{k}_{\parallel}|^2}$. By detuning the cavity photon and exciton energy via the cavity length, the mixing of the  $\mathcal{Y}_1$ exciton with the cavity photon and higher-energy excitons can be modified. This can be understood using the Hopfield approach (see supplementary text) yielding for the inverse effective mass of the polariton
\begin{equation}
\frac{1}{m_n^{(P)}} = \frac{ |U_{0n}(0)|^2}{ m^{(c)}} + 
\sum_{\nu=1}^N   \frac{|U_{\nu n}(0)|^2 }{m^{(Y)}_\nu}, \label{eq:effective_mass}
\end{equation}
The increasing photonic character of $\mathcal{P}_1$ for larger cavities, revealed by the Hopfield coefficients in Fig.~\ref{fig:fig1}(d), manifests as an increased curvature in Fig.~\ref{fig:fig2}, and hence smaller effective mass. Consequently, the effective mass can be tuned over a staggering seven orders of magnitude, from $m^{(Y)} \rightarrow m^{(c)}$. Equation \ref{eq:effective_mass} reveals that one needs to detune the cavity resonance very far from the exciton energy to achieve a polariton mass comparable to the exciton mass. The reason is that for comparable Hopfield coefficients, the term $\propto 1/m^{(c)}$ dominates. As a consequence, when the polariton exhibits even a small photonic character it tends to have an extremely small mass and hence be delocalized over many moir\'{e} unit cells.

\subsection*{Twist Control of Moir\'{e} Exciton Polaritons}
\begin{figure}[t!]
\includegraphics[width=\columnwidth]{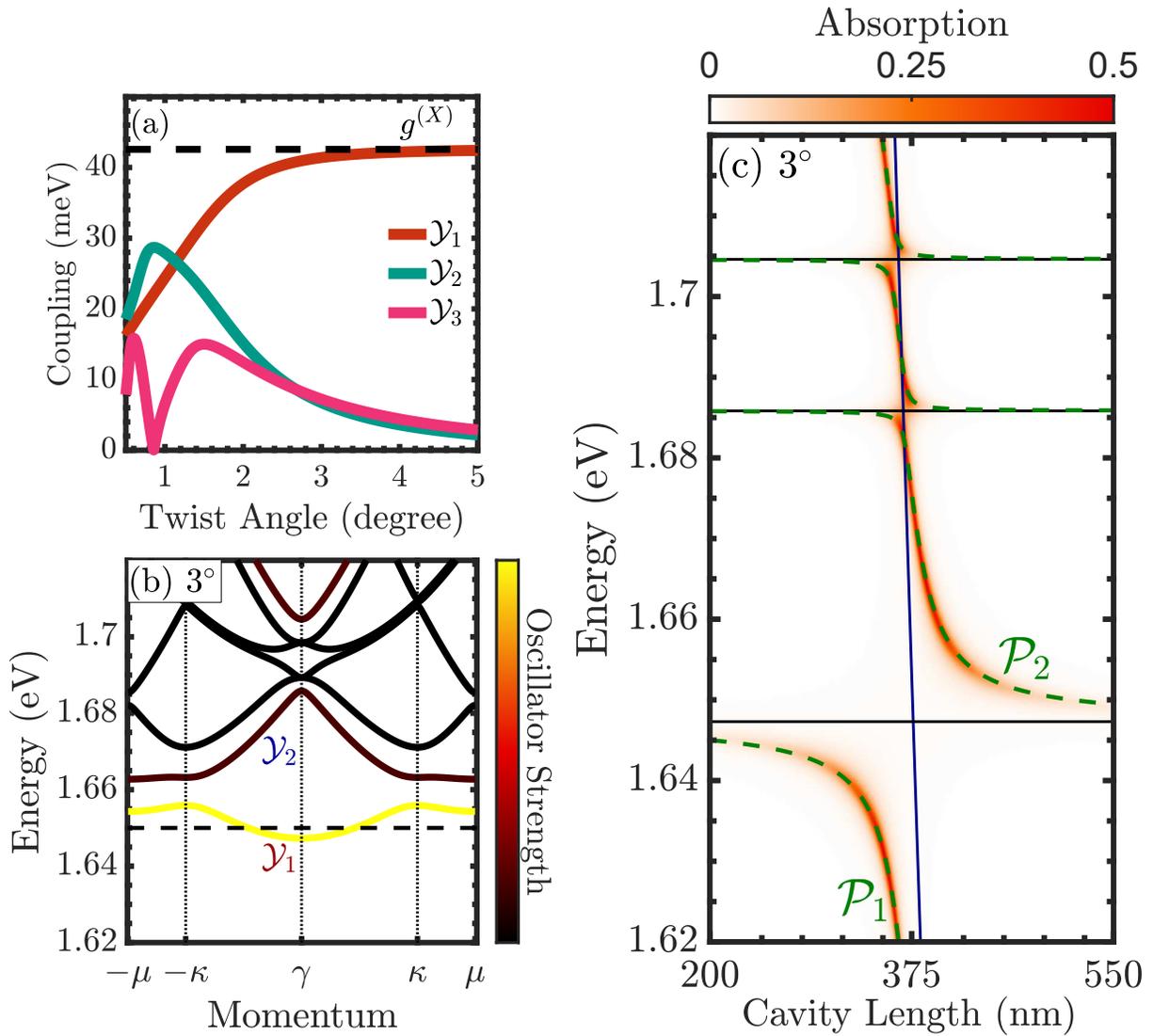}
\caption{\label{fig:fig3} (a)~Coupling strength, $g_\nu$, of the three lowest-energy moir\'{e} excitons as a function of twist angle. The black dashed line shows the radiative coupling of the intralayer exciton in the absence of moir\'{e} effects, $g^{(X)}$. (b)~Moir\'{e} exciton minibands for the intralayer exciton at a $3^{\circ}$ twist angle. Colour coding is proportional to the oscillator strength. (c)~Cavity length dependence of the moir\'{e} polariton bands (dashed-green line) overlaid on the absorption for a twist angle of $3^{\circ}$.}
\end{figure}

The lattice mismatch in \ce{MoSe2}/\ce{WSe2} heterobilayers is very small, and thus the moir\'{e} potential is strongly dependent on the twist angle between the two TMD layers. Here, we investigate the twist-angle dependence of moir\'{e} exciton polaritons. The following effects play an important role: (i) the energy detuning between the cavity photon and each moir\'{e} exciton changes with the twist angle, (ii) a shrinking moir\'{e} supercell (for larger twist angles) leads to delocalized excitons, which can hop to neighbouring cells \cite{brem2020tunable}, and (iii) oscillator strength accumulates with increasing twist angle into the lowest-energy exciton at the expense of all the others. The latter is shown in Fig.~\ref{fig:fig3}(a) for the coupling strength, $g_\nu$, of the three lowest exciton branches. As the twist-angle increases, $g_1$ limits to the untwisted value, $g^{(X)}=42.6$ meV. 

To illustrate the twist-angle dependence, we repeat the cavity length-sweep analysis for an angle of $3^\circ$. At this larger angle, there is significant hopping of excitons between moir\'{e} supercells, which is reflected by a shallow, near-parabolic energy dispersion of the $\mathcal{Y}_1$ exciton, cf. Fig.~\ref{fig:fig3}(b). The significant accumulation of oscillator strength in $\mathcal{Y}_1$ leads to a larger Rabi splitting when compared to the $1^\circ$ case, cf. Fig.~\ref{fig:fig3}(c). The coupling strength $g$ for $\mathcal{Y}_1$ is nearly doubled from $24$ meV at $1^\circ$ to $41$ meV at $3^\circ$, at the expense of all other moir\'{e} excitons, which have a coupling less than $8$ meV at $3^\circ$. The presence of other moir\'{e} excitons is only visible in weaker avoided crossings close to $1.68$ and $1.70$ eV. The large spacing in energy between the three excitons, relative to their interaction strength with light, means that the cavity photon cannot couple them efficiently, resulting in the middle branches having a much steeper photon-like character near the cavity energy relative to what is seen in Fig.~\ref{fig:fig1}(c). 

\begin{figure}[t!]
\includegraphics[width=\columnwidth]{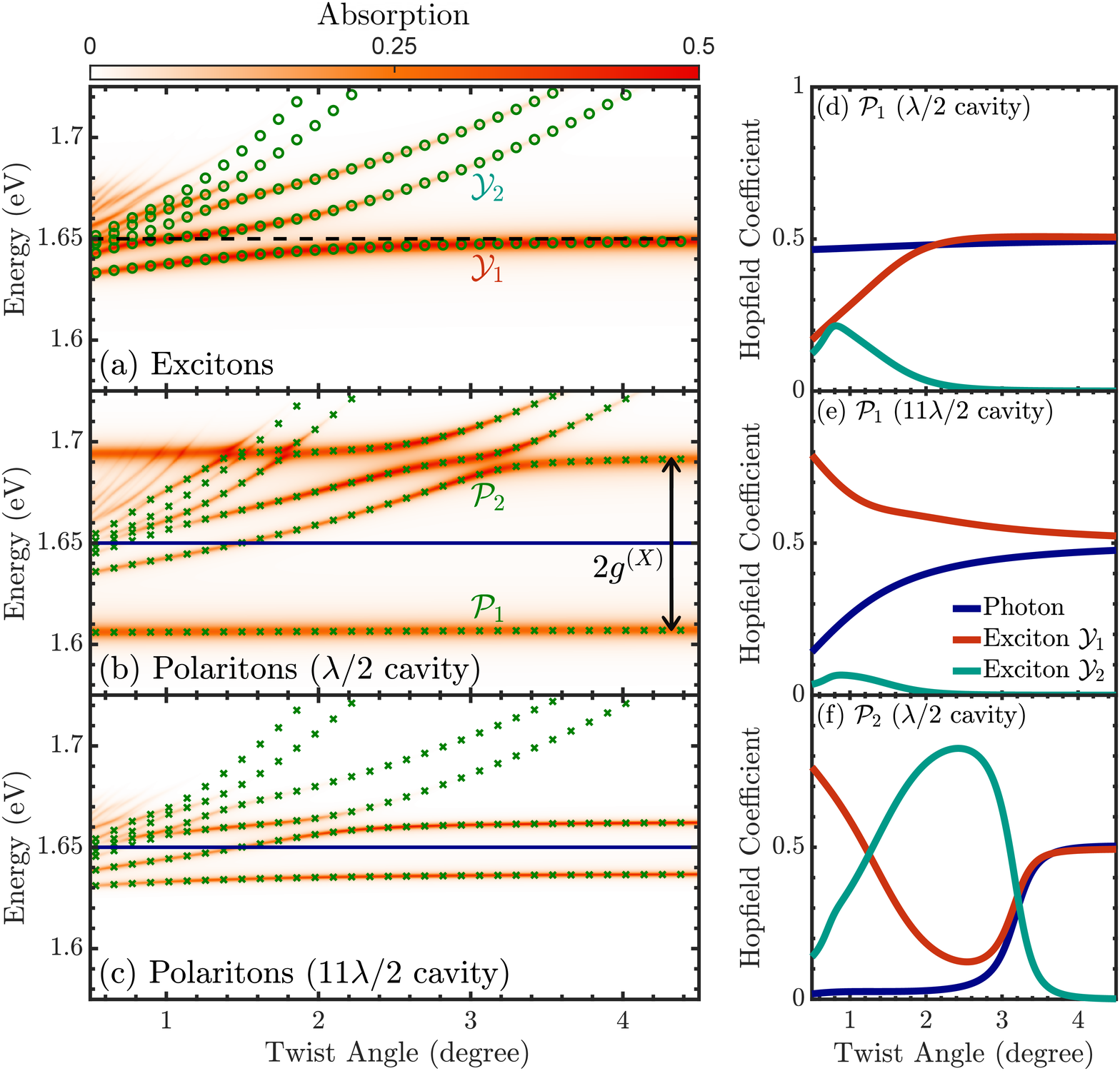}
\caption{\label{fig:fig4} (a)~Twist-angle dependence and absorption of the intralayer moir\'{e} excitons in the absence of a cavity. The exciton energy is shown by the green circles. Twist-angle dependence and absorption of moir\'{e} polaritons (green crosses) in a (b) $\lambda/2=374$ nm and (c) $11\lambda/2=4120$ nm cavity. The cavity mode energy is shown by the flat blue line and is equal to the 1s intralayer exciton energy in the absence of moir\'{e} effects (black-dashed line in (a)). Absolute square of the Hopfield coefficients for the $\mathcal{P}_1$  polariton branch in a (d)~$\lambda/2$ and (e)~$11\lambda/2$ cavity, and (f) the $\mathcal{P}_2$ polariton branch in a  $\lambda/2$ cavity. 
}
\end{figure}

Now, we investigate the polariton dispersion and absorption for a range of twist angles between $0.5^\circ$ and $4.5^\circ$, cf. Fig.~\ref{fig:fig4}. The cavity energy is twist-angle independent (flat blue line), and tuned to the untwisted exciton energy of $1.65$ eV (corresponding to a cavity length of $L=\lambda/2=374$ nm). In contrast, the moir\'{e} exciton energies have a clear twist-angle dependence, cf. Fig.~\ref{fig:fig4}(a). In particular, the $\mathcal{Y}_1$ exciton shifts by $14$ meV as the twist angle is tuned from $0.5^\circ$ to $3^\circ$. We find an abundance of bright excitons, which decrease in number for increasing twist angle \cite{brem2020tunable}. At larger angles, the only notable absorption arises from the lowest $\mathcal{Y}_1$ branch due to the accumulation of oscillator strength. 

Figure~\ref{fig:fig4}(b) shows the polariton dispersion and absorption for the heterobilayer placed within a cavity. The strong coupling leads to the almost completely twist-angle independent $\mathcal{P}_1$ polariton branch that is red-shifted by $g^{(X)}$ relative to the cavity mode energy. The polariton energy shifts less than $1$ meV over the angle range of $0.5^\circ \rightarrow 3^\circ$. At larger twist angles ($\gtrsim 3^\circ$), this is expected due to the oscillator strength accumulating into $\mathcal{Y}_1$ (Fig.~\ref{fig:fig3}(a)), which has a near twist-angle independent energy at these angles (Fig.~\ref{fig:fig4}(a)). This manifests as a twist-angle independent $\mathcal{P}_1$ and $\mathcal{P}_2$ in the strong-coupling regime, and the splitting between the two limits to the untwisted value of $2g^{(X)}=85$ meV (indicated by the black arrow). 
The behaviour of the polariton $\mathcal{P}_1$ at small angles is puzzling, as the analysis of the Hopfield coefficients (Fig.~\ref{fig:fig4}(d)) shows that it has a sizable contribution from $\mathcal{Y}_1$ and $\mathcal{Y}_2$ excitons, which both have a clear twist-angle dependence in the range of small angles. The photonic contribution, on the other hand, is a nearly constant $0.5$ over all twist angles, cf. Fig. \ref{fig:fig4}(d). This leads to the almost twist-angle independent absorption apparent in Fig. \ref{fig:fig4}(b), cf. Eq.~(\ref{eq:absorption}). 

To elucidate further, note that if the coupling strength is decreased by extending the cavity length (see Eq.~(\ref{eq:coupling})), then the $\mathcal{P}_1$ polariton starts to recover a similar twist-angle dependence as the $\mathcal{Y}_1$ exciton, cf. Fig. \ref{fig:fig4}(c) for a $L=11\lambda/2=4120$ nm cavity (energy shift of $5$ meV over the range $0.5^\circ \rightarrow 3^\circ$). The polariton twist-angle dependence can be understood as a compromise between: (i) the exciton-cavity detuning, i.e. blue-shift of $\mathcal{Y}_1$ and hence $\mathcal{P}_1$ with the twist angle (Fig. \ref{fig:fig4}(a)), and (ii) coupling strength, i.e. larger splitting due to accumulating oscillator strength with increasing twist angle (Fig. \ref{fig:fig3}(a)), which leads to a red-shift of $\mathcal{P}_1$ due to the increased Rabi splitting. For the larger coupling, the photon can effectively couple to the $\mathcal{Y}_1$ exciton at all twist angles, and thus possesses a large photonic component of nearly $50\%$ for all angles (Fig.~\ref{fig:fig4}(d)). In contrast, the weaker coupling for the $11\lambda/2$-cavity means that at small angles, $\mathcal{P}_1$ has a larger excitonic component, and consequently inherits a proportion of the twist-angle dependence of the $\mathcal{Y}_1$ exciton, cf. Fig.~\ref{fig:fig4}(e).

The $\mathcal{P}_2$ polariton branch shows a stronger twist-angle dependence than $\mathcal{P}_1$ for both cavity lengths at small angles, shifting about $47$ meV from $0.5^\circ$ to $3^\circ$ for the $\lambda/2$ cavity, before plateauing like $\mathcal{P}_1$. Inspection of the Hopfield coefficients in Fig.~\ref{fig:fig4}(f) reveals that this arises from the large contribution of the $\mathcal{Y}_2$ exciton at intermediate angles. As the twist angle increases above $3^\circ$, the photonic and the $\mathcal{Y}_1$ contribution grows to $0.5$ each, explaining the growing angle-independence. It is interesting to observe a weaker twist-angle dependence of $\mathcal{P}_2$ for the $11\lambda/2$-cavity, cf. Fig.~\ref{fig:fig4}(c). There is now a energy shift of $23$ meV over the range $0.5^\circ \rightarrow 3^\circ$. This illustrates that modifying the cavity length can both reduce and increase the twist-angle dependence of different polariton branches. We observe that all other higher-energy polariton branches tend to a linear twist-angle dependence, following the behaviour of bare excitons for larger angles, where they exhibit a negligible absorption due to the vanishing oscillator strength and large detuning.

\section*{Discussion}
This work sheds light on the dispersion relation and optical response of moir\'{e} exciton polaritons in twisted van der Waals heterobilayers integrated in a Fabry-Perot cavity. Specifically, we have shown that the small spacing between moir\'{e} excitons and the large coupling with light inherent to TMD intralayer excitons leads to a distinct set of moir\'{e} exciton polaritons, which can consist of multiple moir\'{e} excitons due to photon-induced hybridisation. Exploiting the Hopfield approach allows us to characterize energy, absorption, hybridisation and localization of these unique polaritons. The rich twist-angle and cavity-dependence succinctly exemplifies their dual light-matter character. Progress in stacking technologies with controllable small angle increments \cite{zhang2020twist}, and tunable cavity length \cite{dufferwiel2015exciton} could allow for experimental investigation of the predicted intriguing properties of moir\'{e} exciton polaritons.

\section*{Materials and Methods}

\textbf{Moir\'{e} potential:} The moir\'{e} potential describes the twist-angle dependent electrostatic potential felt by charge carriers in one layer induced by neighbouring atoms in the other layer. Using the monolayer eigenstates to expand the moir\'{e} contribution to the bilayer Hamiltonian gives $\hat{H}_{\text{M}} = \sum_{l, \lambda, \vec{k}\vec{k}'} \braket{l \lambda \vec{k}|\hat{V}_{\tilde{l}}|l \lambda \vec{k}'}\hat{a}_{l \lambda \vec{k}}^\dagger\hat{a}_{ l \lambda \vec{k}'}$, where, $l=\{0,1\}$ is the layer index, $\tilde{l}=1-l$, $V_l$ is the effective electrostatic potential created by layer $l$, and $\lambda=\{\text{v},\text{c}\}$ is the band index. To deduce the form of the moir\'{e} matrix element, we employ the model detailed in Ref.~\cite{brem2020tunable}. The Bloch wavefunction for the lth layer is written within a tight-binding approach using an envelope-function approximation around the K-points. The resulting matrix element, $V_{l,\lambda}(\vec{q}) = \braket{ l \lambda \vec{k}+\vec{q}|\hat{V}_{1-l}| l \lambda \vec{k}}$, is then expressed within a two-site approximation, split into a sum of atomic contributions, and finally a first-shell approximation is taken \cite{bistritzer2011moire}: 
\begin{equation*}
V_{l,\lambda}(\vec{q}) = \mathcal{V}_{l,\lambda} \sum_{n=0}^2 \exp\left[i C^n_3 \left(\vec{G}_l^0 + \vec{G}^0_{\tilde{l}}\right)\cdot \vec{d}/2 \right]
\delta_{\vec{q},\vec{g}_n},
\end{equation*}
where $\vec{d}$ is the lateral displacement between the two TMD layers. The vectors $g_n = C^n_3 \left(  \vec{G}^0_{\tilde{l}} - \vec{G}_l^0 \right)$  span the mBZ reciprocal to the periodic moir\'{e} pattern formed from the twisted lattices. The first-shell approximation involves restricting the contribution of reciprocal lattice vectors to $C_3^n\vec{G}^0 \ (n=0,1,2)$. This approximation preserves the $C_3$ symmetry of the moir\'{e} potential, and is the same form as phenomenological formulas used in previous works \cite{wu2017topological,wu2018theory,yu2017moire}.
The potential $\mathcal{V}_l$ quantifies the strength of the potential and is obtained as material-specific external input parameter from first-principles calculations, see Ref.~\citenum{brem2020tunable}.

\textbf{Bilayer Wannier equation:}
Solving the bilayer Wannier equation \cite{kira2006many} gives access to the 1s exciton wavefunction $\Psi(\vec{r})$, and its excitonic binding energy. The screened Coulomb potential is modelled as a generalised Keldysh potential for two aligned and anisotropic slabs~\cite{ovesen2019interlayer}. The heterostructure is \ce{hBN} encapsulated ($\epsilon_{\text{sub}}=4.5$), and the dielectric constants used for the TMD layers are taken from Ref.~\citenum{laturia2018dielectric}. The spectral position of the 1s exciton, $E^{(X)}(0)$, is fixed to $1.65$ eV (1.75 eV) for the \ce{MoSe2}(\ce{WSe2})-based intralayer exciton, which have been extracted from photoluminescence measurements \cite{rivera2015observation,nagler2017interlayer}.

\textbf{Exciton-light coupling:}
The radiative coupling of a bilayer exciton (with no moir\'{e} effects) has the form \cite{kira2006many}
\begin{equation}
    \gammalatin^{(X)} =  \frac{e^2 \left|\Psi(\vec{r}=0)\right|^2 \left|\vec{p}_{cv} \cdot \vec{e}_\sigma  \right|^2}{2 m_e^2 \epsilon_0 n_{bg} c E^{(X)}(0)} ,
\end{equation}
where $n_{bg}$ is the background cavity reflectivity (evaluated at the optical frequency), and $\vec{e}_\sigma$ is a polarization vector of the light. For intralayer configurations, the optical matrix element for transitions near the K point reduces to the usual circular selection rule for monolayers $\vec{p}_{cv}(\vec{q})= \frac{p}{\sqrt{2}} [1 \ i\tau]\delta_{\vec{q},0}$. DFT values for the optical matrix element, $p$, were taken from Ref.~\citenum{ovesen2019interlayer}, and are close to the monolayer value for \ce{MoSe2} \cite{dufferwiel2015exciton}.

\textbf{Calculation details:} Throughout our work, the end mirrors of the cavity are assumed equal and the reflectivity is set to a realistic frequency-independent value $r_m =-0.99$ (corresponding to a quality factor of $158$ and a linewidth of $10.6$ meV for a cavity tuned to $E^{(X)}$) \cite{zhang2021van}. The exciton scattering loss is set to a constant value of $\hbar \Gamma = 1$ meV, which is appropriate for \ce{hBN}-encapsulated monolayers at low temperatures below 100 K \cite{ajayi2017approaching}. We have checked our results by comparing our approach to the classical transfer-matrix method (see supplementary text). We find an excellent agreement for the polariton dispersion and absorption calculated using the two methods.

%BibTeX users: After compilation, comment out the following two lines and paste in
% the generated .bbl file. 

\bibliography{bib}

\bibliographystyle{Science}

\section*{Acknowledgments}
We acknowledge funding from the European Union’s Horizon 2020 research and innovation program under grant agreement No. 881603 (Graphene Flagship), the DFG via SFB 1083 (project B9), and the Knut and Alice Wallenberg Foundation (2014.0226).

% Set equation counter and label form for Supp. material
\renewcommand{\theequation}{S.\arabic{equation}}
\setcounter{equation}{0}
\renewcommand{\thesubsection}{S.\arabic{subsection}}
\setcounter{subsection}{0}
\section*{Supplementary materials}

\subsection{Moir\'{e} Exciton Optics}
The semi-classical exciton-light coupling \cite{haug2009quantum}, (expressed within the minimal-coupling picture and dipole approximation) 
$
\hat{H}_{\text{XL}} = \frac{e}{m_0}\vec{A}_\sigma\cdot \sum_{\vec{k},\vec{q}} \vec{p}_{cv}(\vec{q}) \hat{c}_{\vec{k}+\vec{q}}^\dagger \hat{v}_{\vec{k}} + \text{h.c.},
$
can be written in the in the excitonic basis \cite{kira2006many,brem2020tunable,brem2020microscopic} (focusing only on the 1s exciton)
\begin{equation}
\hat{H}_{\text{XL}} =  A_\sigma \sum_{\vec{Q}} \Omega^{(X)}(\vec{Q}) \hat{X}_{\vec{Q}}^\dagger + \text{h.c.} \quad \text{with}   \quad
\Omega^{(X)}(\vec{Q}) = \frac{e}{m_0} \vec{p}_{cv}(\vec{Q})\cdot\vec{e}_\sigma \sum_{\vec{k}} \psi(\vec{k}),
\end{equation}
where $ \vec{p}_{cv}(\vec{Q})$ is the optical matrix element, and $\vec{e}_\sigma$ the polarization vector of the light. Switching now to the moir\'{e} exciton basis, the light-exciton Hamiltonian reads \cite{brem2020tunable,brem2020microscopic}
\begin{equation}
\hat{H}_{\text{YL}} = A_\sigma \sum_\nu \Omega^{(Y)}_{\nu} \hat{Y}_\nu^\dagger + \text{h.c.} 
 \quad \text{with}  \quad 
  \Omega^{(Y)}_{\nu} = \sum_{i,j} \mathcal{C}_{\nu,ij}(\vec{Q}=0) \Omega^{(X)}\left(i\vec{g}_1+j\vec{g}_2\right),
\end{equation}
where only states within the light cone, $\vec{Q}=0$, interact with light.
The radiative coupling for moir\'{e} excitons can be expressed as
\begin{equation}
\gammalatin_{\nu} = \frac{| \Omega^{(Y)}_{\nu}|^2}{2 \epsilon_0 n_{bg} c \hbar \omega },
\end{equation}
where $n_{bg}$ is the background cavity reflectivity (evaluated at optical frequency). Typically, the frequency dependence is ignored \cite{kira2006many}, and $\hbar \omega$ is set to the exciton energy (i.e. Eq.~(5) in the main text). If we compare to the radiative coupling of the bilayer without moir\'{e} effects $
    \gammalatin^{(X)} =  \frac{ |\Omega^{(X)}|^2}{2 \epsilon_0 n_{bg} c \hbar \omega}, 
$
we find
\begin{equation}
 \gammalatin_{\nu}^{(Y)} = 
\left|\mathcal{C}_{\nu,i=j=0}(\vec{Q}=0)\right|^2  \gammalatin^{(X)}, \label{eq:oscillator_strength}
\end{equation}
revealing that the oscillator strength (which is proportional to the radiative coupling) is conserved due to the normalisation of the moir\'{e} wavefunctions, i.e. $\sum_{\nu} \left|\mathcal{C}_{\nu,i=j=0}(\vec{Q}=0)\right|^2 = 1$.

\subsubsection*{Hopfield Method}
The fully quantized exciton-light interaction for the 1s exciton in the absence of moir\'{e} effects can be written as 
\begin{equation}
\hat{H}_{XL} = g^{(X)} \hat{X}^\dagger \hat{b} + \text{h.c} \quad \text{with}\quad
g^{(X)} = \hbar \sqrt{\left( \frac{1+|r_m|}{|r_m|} \right) \frac{\gammalatin^{(X)}}{\tau}}.
\end{equation}
The coupling $g^{(X)}$ is calculated by solving Maxwell's equations for a 2D excitonic layer centered in a Fabry-Perot cavity and approximating the cavity to be tuned close to the exciton energy \cite{kavokin2003thin}. Despite the approximation, we find excellent agreement with the T-matrix method (see supplementary section \ref{sec:T-matrix}), even away from the exciton energy. Transformation into the moir\'{e} exciton basis is simple assuming the dipole approximation and neglecting the $Q$-dependence of $g^{(X)}$. Using Eq.~(\ref{eq:oscillator_strength}), this results in the Eq.~(2) of the main text.

\subsection{Transfer Matrix Method} \label{sec:T-matrix}
An alternative framework of a more classical flavour compared to the Hopfield method is the transfer(T)-matrix method. Here, the heterobilayer is modelled as a two-dimensional sheet current and a microscopic or phenomenological model is used to find the response function. This sort of technique has been applied to model the optics of graphene \cite{zhan2013transfer} and monolayer TMDs \cite{vasilevskiy2015exciton}, and has a long history in cavity polaritonics \cite{savona1995quantum}. The advantage of this method is (i) its generality, for instance an arbitrary number of layers can be modelled, (ii) the ease in which experimentally relevant quantities such as reflection and absorption can be calculated, and (iii) the separation into distinct material (calculating the response function) and optics (solving Maxwell's equations) problems. In contrast, the Hopfield method is dependent on an analytical expression for the light-exciton coupling $g$ (or it can be treated as an external parameter), and the calculation of optical spectra requires a more complicated quantum Langevin approach (see Supplementary section \ref{sec:langevin}). A disadvantage of the T-matrix method is that there is no access to the Hopfield coefficients, and it is difficult to extend to nonlinear physics such as polariton-polariton interaction. The two methods are equivalent in the limit of linear optics \cite{kavokin2003thin}, and we find excellent agreement for the polariton dispersion and absorption calculated using the two methods, see Fig.~\ref{fig:fig_supp_1}(a). 

\subsection{Linear Optical Spectra of Polaritons} \label{sec:langevin}
Cavity polaritons can couple to the outside universe via the non-zero transmission of the cavity mirrors. This can be modelled using the Heisenberg-Langevin equations along with the input-output relations \cite{collett1984squeezing}, which works well in the limit of high-Q cavities (i.e. near perfect end mirrors, $|r_m| \sim 1$). Using this approach, we quantize separately the internal cavity mode and the external radiation fields, which are weakly coupled via the end mirrors. This leads to a consistent description of both the radiative decay rate, and the coupling of the polaritons to input and output fields. The interaction between the cavity and external radiation modes is assumed to have a simple quadratic form and the rotating-wave approximation is used. It is also relatively straightforward to add exciton loss in a similar fashion. Each exciton is coupled to a phonon bath, described by a phenomenological value $\hbar \Gamma_\nu$. With these assumptions and using $\hat{b} = \sum_n U_{0n} \hat{\mathcal{P}}_n$ and $\hat{Y}_\nu = \sum_n U_{\nu n} \hat{\mathcal{P}}_n$, the total Hamiltonian can be written in the polariton basis as
\begin{equation}
\begin{split}
    \hat{H} &= \sum_{n=0}^N E_n^{(Y)} \hat{\mathcal{P}}^\dagger_n \hat{\mathcal{P}}_n + \sum_{j=L,R}  \int d\omega \  \hbar \omega \hat{\mathcal{B}}_{j,\omega}^\dagger\hat{\mathcal{B}}_{j,\omega} + 
    \sum_{\nu=1}^N \int d\omega \  \hbar \omega \hat{\mathcal{D}}_{\nu,\omega}^\dagger\hat{\mathcal{D}}_{\nu,\omega}
    \\
    &+i\hbar \sum_{j=L,R} \sum_{n=0}^N \int d\omega \ \sqrt{2\kappa_j} \left( U_{0n}\hat{\mathcal{B}}_{j,\omega}^\dagger\hat{\mathcal{P}}_n - U_{0n}^*\hat{\mathcal{B}}_{j,\omega}\hat{\mathcal{P}}_n^\dagger \right)
    \\ &+ i \hbar \sum_{\nu=1}^N \int d\omega \ \sqrt{2\Gamma_\nu} \sum_{n=0}^N\left[U_{\nu n}\hat{\mathcal{D}}^\dagger_{\nu,\omega}\hat{\mathcal{P}}_n -  U_{\nu n}^*\hat{\mathcal{D}}_{\nu,\omega}\hat{\mathcal{P}}_n^\dagger \right], 
\end{split}
\end{equation}
where the left-hand and right-hand external modes are described by the field operators $\hat{\mathcal{B}}_{L, \omega}$ and $\hat{\mathcal{B}}_{R, \omega}$ respectively, and the phonon bath operator is described by $\mathcal{D}_{\mu, \omega}$. The coupling parameters $\kappa_L$, $\kappa_R$ and $\Gamma_\nu$ are frequency dependent in general, but we take the Markov approximation. The presence of the Hopfield coefficients reflects the fact that only the photonic (excitonic) part of the polariton couples to the external radiation field (phonon baths).

The dynamics for the polariton and bath operators can be found from the Heisenberg equation of motion. They can be solved by the usual prescription of first finding the formal solution of the bath operators in terms of an initial time $t_0$, and then substituting into the equation of motion for the polaritons. Thanks to the Markov approximation, a simple first-order differential equation in time is found to govern the polariton dynamics. In particular, if we assume the cavity to be only driven from one port, and that the phonon coupling term has a constant value for each moir\'{e} exciton, we find
\begin{equation}
\begin{split}
\frac{d}{dt}\mathcal{P}_n(t) &= \left(-i\frac{E_n^{(Y)}}{\hbar} 
+\Gamma \right)\mathcal{P}_n(t)  +U_{0n}^* \sqrt{2\kappa_L} a_{\text{in}}(t) \\
&-U_{0 n}^*(\kappa_L+\kappa_R-\Gamma ) \sum_{m=0}^N  U_{0 m} \mathcal{P}_m(t), \label{eq:polariton_EOM}
\end{split}
\end{equation}
where $a_{\text{in}}(t)=-\int d\omega \ \hat{\mathcal{B}}_{j,\omega}(t_0)\exp[-i\omega(t-t_0)]$ is the input field and contains contributions from both vacuum quantum noise and a classical driving field \cite{collett1984squeezing}. As we are only interested in the mean field (large photon number), we drop all fluctuation terms. Boundary conditions at the mirrors are given by the input-output relations, allowing the input field to be related to the outgoing fields in each port \cite{collett1984squeezing}
\begin{equation}
a_{\text{ref}}(t) = -a_{\text{in}}(t) + \sqrt{2\kappa_L} \sum_{n} U_{0n} \hat{\mathcal{P}}_n(t), \quad
a_{\text{trans}}(t) =  \sqrt{2\kappa_L} \sum_{n} U_{0n} \hat{\mathcal{P}}_n(t). \label{eq:input_output}
\end{equation}
Then, Eqs.~(\ref{eq:polariton_EOM}) and (\ref{eq:input_output}) can be Fourier transformed and combined to give the reflection and transmission coefficients
\begin{equation}
\begin{split}
r(\omega) &= \frac{a_{\text{ref}}(\omega)}{a_{\text{in}}(t)} =  \frac{-1 + (\kappa_L - \kappa_R+\Gamma)\Pi(\omega)}{1+(\kappa_L + \kappa_R-\Gamma)\Pi(\omega)}
\\[6pt]
it(\omega) &= \frac{a_{\text{trans}}(\omega)}{a_{\text{in}}(t)} =  \frac{\sqrt{2\kappa_R} \sqrt{2\kappa_L} \Pi(\omega)}{1+(\kappa_L + \kappa_R-\Gamma) \Pi(\omega)}, 
\end{split} \label{eq:langevin_equations}
\end{equation}
with $ \Pi(\omega) =  \sum_{n=0}^N \frac{|U_{0n}|^2}{i\left( E_n^{(Y)}/\hbar-\omega \right)+\Gamma}$. It can be checked that Eqs.~(\ref{eq:langevin_equations}) correctly describes a bare Fabry-Perot cavity in the limit of a pure photonic mode, i.e. $|U_{0n}|=1$. 

The only task remaining is to determine the coupling constants $\kappa_L$ and $\kappa_R$. By using the exact analytical expression for a lossless and symmetric bare Fabry-Perot cavity, it is possible to expand for frequencies close to the cavity frequency and a high-quality cavity ($T_m\approx 0$), and compare to equations \ref{eq:langevin_equations} to find $\kappa_L = \kappa_R = c T_m/(4L)$.

It is straightforward to find an expression for the absorption from energy conservation $A(\omega)=1-|r(\omega)|^2-|it(\omega)|^2$. Taking a symmetric cavity and using Eqs.~(\ref{eq:langevin_equations}) gives
\begin{equation}
    A(\omega) = \frac{4\kappa(\Re[\Pi(\omega)]-\Gamma|\Pi(\omega)|^2)}{|1+(2\kappa-\Gamma)\Pi(\omega)|^2}. \label{eq:absorption_2}
\end{equation}
It is instructive to make further simplifications to find an approximate analytical expression for the absorption. Inspired by the similar derivation of the Elliot formula \cite{brem2020microscopic,kira2006many}, we expand the numerator and denominator and neglect interference effects. This is justified for energetically well-separated moir\'{e} excitons relative to the scattering rate $\hbar\Gamma$. This returns Eq.~(3) of the main text,
\begin{equation}
    A(\omega)= \sum_{n=0}^N \frac{4 \tilde{\gammalatin}_n  \tilde{\Gamma}_n}{( \omega- E_n^{(Y)}/\hbar)^2 + (2 \tilde{\gammalatin}_n + \tilde{\Gamma}_n)^2}, \label{eq:absorption_3}
\end{equation}
which has a remarkable similarity with the conventional Elliot formula. While it does not include light-matter fully self-consistently like Eq.~(\ref{eq:absorption_2}), we find  an excellent agreement with the T-matrix method. This is demonstrated in Fig.~\ref{fig:fig_supp_1}(a).

\subsection{Polaritonic Critical Coupling Condition}
Here we discuss the conditions for peak absorption in Eq.~(\ref{eq:absorption_3}) (Eq.~(3) of the main text). Focusing only on the polariton $n$ and evaluating the equation at $\hbar \omega = E_n^{(Y)}$ gives 
\begin{equation}
A=  \frac{4 \tilde{\gammalatin}_n  \tilde{\Gamma}_n}{(2 \tilde{\gammalatin}_n + \tilde{\Gamma}_n)^2}. 
\end{equation}
The maximum absorption of $0.5$ is reached at the condition $2\tilde{\gammalatin}_n = \tilde{\Gamma}_n$. The value of the Hopfield coefficients that satisfy this condition depend on the ratio between the exciton scattering rate, $\Gamma$, and the cavity decay rate $\kappa$. For the specific case where the non-radiative decay rate, $\Gamma$, is balanced by the total radiative decay through both ports of the bare cavity, $2\kappa$, the maximum absorption is achieved for an equal light-matter contribution to the polariton, i.e. $|U_{0n}|^2=0.5$. For the parameters chosen throughout this work, $\Gamma=1$ meV and $|r_m|=0.99$, the critical coupling condition for the $\mathcal{P}_1$ polariton at a $1^\circ$ twist angle is found at the cavity length $L_{\text{max}}=364$ nm (black-dotted line), which corresponds to a Hopfield coefficient of $|U_{0n}|^2=0.16$, cf. Fig.~\ref{fig:fig_supp_1}(a). For more insight, we show in Fig.~\ref{fig:fig_supp_1}(b) a plot of the effective decay rates, as well as the corresponding bare-cavity decay rate and exciton scattering rate, as a function of the cavity length. We can observe that the critical coupling condition occurs (crossing point of the blue- and red-solid line) at the peak absorption cavity length of $L_{\text{max}}$ (black-dotted line).

\subsection{Polariton Group Velocity and Effective Mass}
To derive an expression for the polariton group velocity for a general number of excitons and photons, we start by evaluating the commutator relation $[\hat{\mathcal{P}}_n,\hat{H}]$ twice: once for the Hamiltonian expressed in the diagonal form with polariton operators, and then expressed in the original form with exciton and photon operators. This gives two equations
\begin{equation}
\begin{split}
\left(E_n^{(P)}(\vec{k}_\parallel) - E_c(\vec{k}_\parallel) \right) U_{0n}(\vec{k}_\parallel) &= \sum_{\nu=1}^N  g_\nu(\vec{k}_\parallel)  U_{\nu n}(\vec{k}_\parallel) 
\\
\left(E_n^{(P)}(\vec{k}_\parallel) -  E_\nu^{(Y)}(\vec{k}_\parallel) \right) U_{\nu n}(\vec{k}_\parallel) &=    g_\nu(\vec{k}_\parallel) U_{0n}(\vec{k}_\parallel),
\end{split}
\end{equation}
where $E_c$ is the cavity photon energy. The second equation can be plugged into the first equation resulting in
\begin{equation}
    \left(E_n^{(P)}(\vec{k}_\parallel) - E_c(\vec{k}_\parallel)\right)  = \sum_{\nu=1}^N  \frac{  g_\nu^2(\vec{k}_\parallel)   }{E_n^{(P)}(\vec{k}_\parallel) -   E_\nu^{(Y)}(\vec{k}_\parallel)   }.
\end{equation}
This expression can be differentiated with respect to in-plane momentum, and rearranged to give the group velocity for the $n$th polariton
\begin{equation}
v_n(\vec{k}_\parallel) =   \left|U_{0n}(\vec{k}_\parallel)\right|^2 v_c(\vec{k}_\parallel) + \sum_{\nu=1}^N   \left|U_{\nu n}(\vec{k}_\parallel)\right|^2 v_\nu(\vec{k}_\parallel) +  2 U_{0n}(\vec{k}_\parallel)\sum_{\nu=1}^N  U_{\nu n}^*(\vec{k}_\parallel) \partial_{\vec{k}_\parallel} g_\nu(\vec{k}_\parallel)/\hbar \label{eq:group_velocity}
\end{equation}
This is a generalisation of the expression for the group velocity found in Ref.~\cite{freixanet2000plane} for a single type of exciton and photon. For momenta close to $\vec{k}_\parallel=0$, the first derivative of the coupling will be small and third term can be ignored. In fact, this is exact for TM-polarized cavity modes, where the coupling is constant with in-plane momentum \cite{panzarini1999cavity}. For an isotropic dispersion, which is valid at the $\gamma$ point of the exciton dispersion, the polariton effective mass can then be found from the group velocity using 
\begin{equation}
m_n = \frac{\hbar}{\partial_{\vec{k}_\parallel} v_n(\vec{k}_\parallel)|_{\vec{k}_\parallel=0}},
\end{equation}
which, combined with equation \ref{eq:group_velocity}, gives Eq.~(4) of the main text. Note that we can ignore the derivatives of the Hopfield coefficients as the first derivative vanishes at $\vec{k}_\parallel=0$, reflecting the even symmetry at this point.

%%%%%%%%%%%%%
%% Figures %%
%%%%%%%%%%%%%

\clearpage

% Figure for supplementary material
\renewcommand{\thefigure}{S\arabic{figure}} % Set figure label
\setcounter{figure}{0}

\begin{figure}[t!]
\includegraphics[width=\columnwidth]{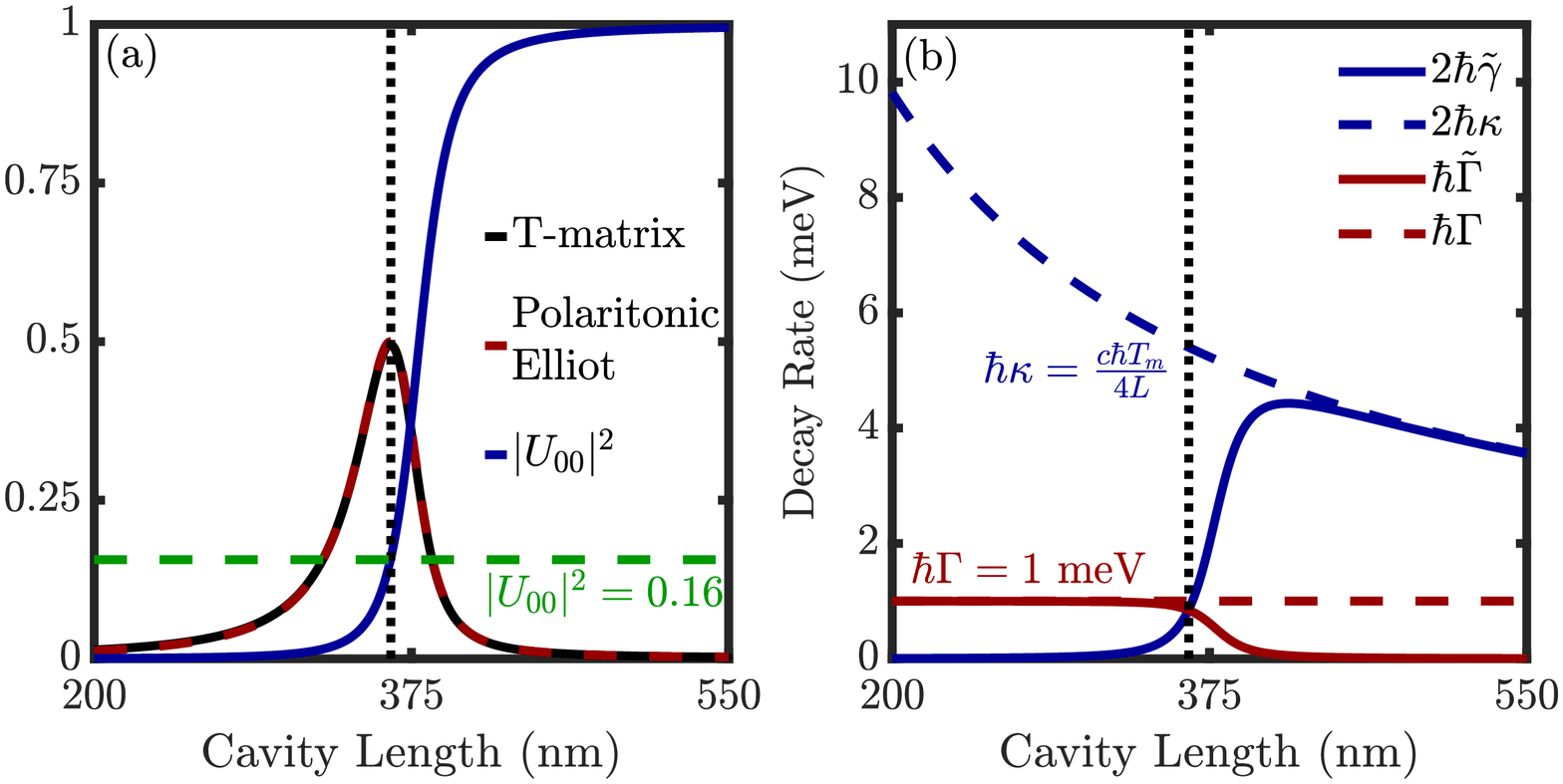}
\caption{\label{fig:fig_supp_1} (a)~Resonant absorption for the $\mathcal{P}_1$ polariton  at a $1^\circ$ twist angle calculated with the T-matrix method (black line) and the polaritonic Elliot formula (Eq.~(\ref{eq:absorption_3}), dashed-red line). The Hopfield coefficient describing the photonic contribution ($U_{00}$, blue line) is also shown. The peak absorption of $0.5$ is indicated by the black-dotted line, and the corresponding value of the Hopfield coefficient that gives this condition is indicated by the dashed-green line. (b) Comparison of twice the effective radiative coupling (blue line), and the effective scattering rate (red line), against cavity length. The bare-cavity decay rate and exciton scattering rate are shown by the blue- and red-dashed lines respectively. The peak absorption (black-dotted line) is given by the polaritonic critical coupling condition of $2\tilde{\gammalatin}=\tilde{\Gamma}$.
}
\end{figure}

\end{document}